\documentclass{optica-article}

\journal{opticajournal} 

\articletype{Research Article}

\usepackage{lineno}

\begin{document}

\title{Lung Diseases Image Segmentation using Faster R-CNNs}

\author{Mihir Jain,\authormark{1} }

\address{\authormark{1}Stevens Institute of Technology\\}

\email{\authormark{*}mjain24@stevens.edu} 


\begin{abstract*} 
 Lung diseases are the number one cause of child mortality across the developing world, about  half of the global pneumonia deaths occur in India 370k according to Indian academy of  pediatrics and national Health profile 2016. Timely diagnosis could significantly reduce the  mortality level. This paper uses a low density neural network structure to avoid topological lag which comes  with network depth and breadth. The neural network integrates the parameters into a feature  pyramid network, extracting data to avoid losses. Softening Non maximal suppression is applied on regional proposals created by RPN. All function information is filtered and a  model to achieve a high level of detection is obtained. For performance evaluation the trained model is validated with randomly chest x-ray images  taken from the same data set, we then compute the confusion matrix to obtain the accuracy,  precision, sensitivity, specificity values. All the loss functions involved in training and testing  model is shown by total loss. The loss function decreases when the training is longer. The regional proposal loss gives the  performance of the model during the training phase and determines the quality of the regional  proposals. The classification loss demonstrated the loss function during the classification  phase and determines the quality of classification.

\end{abstract*}

\section{Introduction and Motivation }
According to Indian Academy of Pediatrics and National Health Profile 2016, [1] Fifty percent  of pneumonia deaths occur in India which means approximately 370k children die of  pneumonia annually in India. Moreover Pneumonia is the number one killer of children causing  18\% of all child mortality in the world. This harsh number is still minuscule compared to  mortality due to other chest lung diseases. 
Timely diagnosis and treatment could reduce the mortality level. The conventional X-ray  images are slow[2] making human evaluation inept. Computer aided diagnosis can enhance  productivity and lead to timely treatment potentially saving millions of lives. 
The size, shape and position of pneumonia can vary greatly[3]. The outline is very vague and  blurry which leads to great difficulty for detection and enhancing the accuracy of detection is  a major research problem. 
At present detection algorithms include two stage region detectors and object classification  such as Faster R-CNN [4] and one stage region detectors and object classification such as  YOLO [5] and SSD [6]. 
In latter object classification and bounding box regression are done directly without using pre  generated region proposals. While in two stage detectors there is generation of region proposals  and then object classification for each region proposal. Though faster than two stage detectors  one stage detectors are less accurate. 
Treatment and Diagnosis needs high accuracy and thus two stage detectors and classifiers have  advantage. 
However, there are still problems with the present image classification models such as  Xception [7] and VGG [8]. They need a large network depth leading to prolonged training time  and large downsampling that leads to the target position and semantic information being lost.

\subsection{Project Statement }

Object Detection is one of the most promising fields to expand global healthcare services  especially in lower income countries. Automated algorithms and tools have the potential to  support workflow, improve efficiency, and reduce errors. The Aim of this paper is to address  common classification problem 
  
The ideal aim would be to assess images using a deep learning feature map, since such a map  has a big receptive field which makes the size of the region in the input that produces the feature  anchor also large . However, with deep maps, it reduces the object-edge resolution, which  reduces the assessment accuracy of the regression curve. So after continuous downsampling in  such maps, the semantic features of the targets which are small disappear in the layers; the large target is also partially lost, and the edge will move, which is not favorable for accurate  detection of the target. 
Conventionally, to optimize a network, such as X-Ception, is to extend the width or depth [9],  but this method creates huge numbers of parameters, which leads to high variance in the model,  and needs large amounts of data to train. 

\subsection{Organisation of Report}
This paper uses a low density neural network structure to avoid topological lag which comes
with network depth and breadth. The neural network integrates the parameters into a [FPN]
feature pyramid network [10], extracting data to avoid losses. Softening-Non maximal
suppression [S-NMS] [11] is applied on regional proposals created by RPN. all functional
information is filtered and a model to achieve high level detection is obtained. This paper uses
Faster RCNN using ResNet.

\section{Background Material}

\subsection{Conceptual Overview}
The reason why one cannot proceed with object detection by making a neural network
(convolutional) followed by a layer is that the length of the output layer is not fixed as the
number of occurrences of the object of interest cant be fixed. One might think to take different
regions of interest and then filter the presence of objects within those regions. The problem
with this approach is that objects of interest could have different aspect ratios and different
locations within the image. Hence a huge number of regions and this could computationally
blow up. Thus arises the need for algorithms like R-CNN, YOLO, SSD.
Conventionally object detection models used 3 steps.
The first step involved generating large number of region proposals. Region proposals are
basically the useful part of image which might have the objects to be detected in them.
Algorithms such as selective search and EdgeBoxes generate region proposals.
Then from each of these region proposals a feature vector is extracted using image descriptors
such as histogram of object gradients. This feature vector is critical for the model to work
correctly these vectors should decribe an object even if it varies in scale or transition.
The feature vector then gives region proposals to the object classes. But as the object classes
increase the complexity for such models vary greatly with classes. After feature extraction the
method which is used for classifying the region proposals are like support vector machine.

\subsection{YOLO}
YOLO (You only look once) was proposed by Redmon in 2016.
As stated in the original proposal [5] “Compared to other region proposal classification
networks (fast RCNN) which perform detection on various region proposals and thus end up
performing prediction multiple times for various regions in a image, Yolo architecture is more
like CNN (convolutional neural network) and passes the image (nxn) once through and output
is (mxm) prediction. This the architecture is splitting the input image in mxm grid and for each
grid generation 2 bounding boxes and class probabilities for those bounding boxes”
The Biggest advantage of the model is speed (45 frames per second) and an even faster version
155 fps which is less accurate due to smaller architecture.
But YOLO imposes strong constraints on bounding box predictions as it treats them as a
regression problem . This limits the number of small objects that a model can predict. Thus the
model struggles with small objects that appear in groups and thus accuracy is low.

\subsection{SSD}
SSD (Single Shot Detection) was proposed by Liu [6] takes only one multiscale feature map to
detect independently, It is significantly faster in speed and accuracy object detection. The
comparison between speed and accuracy of different object detection models on
VOC2007 [12]

SDD300 : 59 FPS with mAP 74.3\%
SSD500 : 22FPS with mAP 76.9\%
Faster RCNN : 7 FPS with mAP 73.4\%
YOLO : 45 FPS with mAP 63.4\%

SSD has a base VGG-16 network followed by multibox layers.
The high speed and accuracy is because it eliminates the bounding box proposals like one used
in R-CNN and filters with different sizes, and aspect ratio for object detection.
The spatial resolution is reduced which makes it unable to locate small targets reducing
accuracy.

\subsection{Region Proposals}
Region Proposals or Region of Interests on given an input image find all the possible places
where the object can be located. The output given is a list of bounding boxes of likely positions
of the objects.

\subsection{RoI Polling}
RoI pooling is a layer neural network used for object detection. It achieves speedup for both
training and inference while maintaining high accuracy.
Consider we need to perform RoI pooling on the map below for one region of interest and an output of size 2x2.Also say we have a region proposal (x,y,h,w coordinates).By dividing into 2x2 size.Note that the size of RoI need not be perfectly divisible by number of pooling sections The
max values are also the output from RoI pooling layer.

\subsection{RCNN}
R-CNN was proposed by Girschick in 2014 [13] to solve the problem of selecting a high
number of regions. He proposed a method to selectively approach search to extract only regions
from the image which are data giving instead of going through all the regions.
This improved the speed of training and the accuracy on VOC2010 dataset the mAP from
35.1\% to 53.7\%. Compared to conventional model the R-CNN model can detect 80 different types of objects in
images extracting features on the basis of convolutional neural network. Other than this
everything is same from the traditional model.
RCNN contatins 3 steps.
The first module generates 2k region proposals using selective search algorithm. Then after
augmentation the second module extracts feature vector of each region proposal proportional
to length of 4,096. The third module uses a pre-trained SVM algorithm to classify region
proposal as either one of the object classes or the background.

The issue with R-CNN is that it still takes a huge time to train the model network as one has to
classify region proposals on each image.
Also it can’t be implemented in real time. The selective search approach is a fixed algorithm
and therefore no learning happens at the stage which could lead to bad candidate region
proposals.

\subsection{Fast RCNN}
This is an object detector which was also developed by Girshick, It overcomes some of the
issues of R-CNN.
- He Proposed a new layer called ROI pooling that extracts equal-length feature vector
from all region of interests (i.e. proposals) in the same image.
- Compared to R-CNN which has multiple steps, Fast R-CNN builds the network in a
single step.
- Fast R-CNN shares convolutional layer calculations across all region of interests rather
than doing calculations for each of proposal singularly. Using the ROI Pooling layer
making the Fast R-CNN faster than R-CNN.
The feature map from the last convolutional layer is fed to an ROI Pooling layer. The reason
is to extract a fixed length feature from each ROI.
ROI Pooling works by splitting each region proposal into a grid of cells. The max pooling operation is applied to each grid cell to return a single value.
After which the extracted feature vector is passed to some neural layers. The output of which
is splitted in 2 branches softmax layer to predict the scores and FC (neural network) layer to
predict the bounding boxes for detected object.

\begin{figure}[h]
    \centering
    \includegraphics[width=0.7\textwidth]{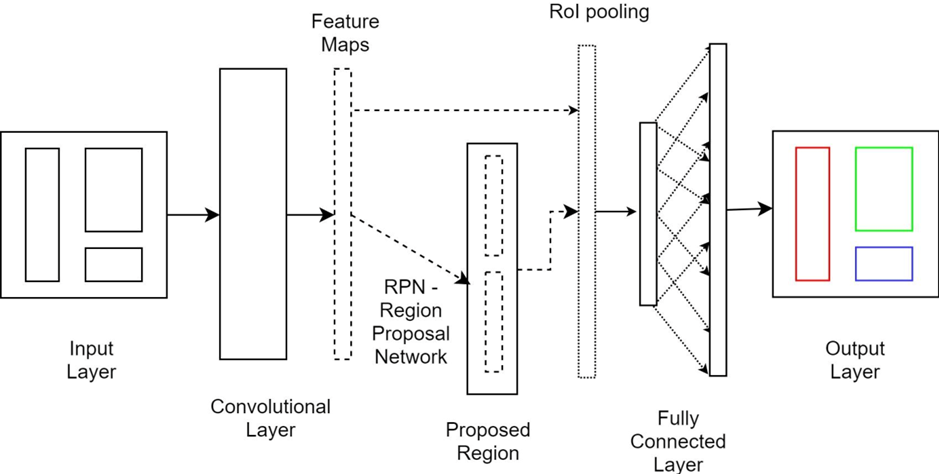}
    \caption{Fast RCNN Architecture}
    \label{fig:fast_Rcnn}
\end{figure}

This leads to decrease in time consumption as fast rcnn shares the computationals across
proposals while rcnn computes for each proposal. Rcnn also takes one single RoI from input
image which if say model need 256 proposals it would take from 256 images while the fast
rcnn could potentially take 256 RoI from 4 images about 64 per image leading to a reduced
time span. Though using multiple layers on the same image reduces the accuracy of the model
as all regions start to correlate.
Even though fast rcnn is better compared to rcnn on time it has its problems as it depends on
selective search to generate region proposals which cannot be customized for a specific object
detection dataset.

\subsection{Faster RCNN}
Selective search [14] is a slow and time consuming process to overcome which in 2015 Ren
[4] came up with the Faster R-CNN model, an object detection algorithm that doesn’t need to
search and lets the network learn the region proposal.It is an extension of Fast R-CNN due to region propsoal network RPN
Region proposal network is a convolutional neural network that generates proposals with
various augmentations. It tells the model network for where to look.

Instead of using various images at different shape or sizes the paper introduced anchor boxes.
An anchor box is a reference box of specific scale and ratio. Multiple reference boxes leads to
multiple share and sizes for binary region. Each region is then mapped to reference anchor box
for detecting at different scales.
The computations are shared across the RPN and Fast-RCNN proposed above to reduce the
computational time.
The architecture of Faster R-CNN  :- consists of two parts
RPN : For detecting region proposals \& Fast R-CNN for object detection in the proposed
regions.

It works as follows
RPN generates region proposals. Then from all region proposals a fixed length feature vector
is extracted using RoI Pooling layer. Then extracted features are classified using Fast r-cnn.
Then class scores of detected objects in addition to their boxes are returned.
Both r-cnn models before faster rcnn depend on selective search algorithms to generate region
proposals. Each proposal is fed to a pre trained CNN while faster rcnn uses region proposals
network to produce region proposals.Thus region proposals are produced using a network
which means they can be trained for specific tasks in object detection. Also as they are trained
now the model can be used on customised dataset which produces better proposals than
selective search or EdgeBoxes.By sharing convolutional layers, the RPN is merged with Fast-rcnn to a single unified network
to do training only once.
RPN works on the output feature map of the convolution layer shared with the Fast-rcnn a
sliding window passes over each of the feature map that leads to generation of region proposal.
Each proposal is characterized by score given by a reference box called the anchor box. The
anchor box has two parameters Scale and aspect ratio.

k regions are produced from each region proposal where k regions varies in scale or size. As
anchor boxes vary in different sizes and scale they are able to get a non changing scale object
detector as a single image at a single scale is used. Multi scale anchor boxes are to the benefit
of the model.
From region proposal a feature vector is extracted and fed to two layers. The first is a binary
classification that generates object score for each proposal and second returns the bounding
box of region. First layer has two outputs whether the region is object that had to be detected or
the background.
The layer outputs two elements if the first is 1 and second is 0 then region is classified as
background wherelse if the first is 0 and second is 1 it is the object.
For RPN training each anchor is also given a score based on IoU which is later discussed is the
ratio of intersection of area between the anchor box and ground truth to the union of the same
boxes. IoU increases as the boxes come closer to each other.
The image is used as an input to a network which outputs a convolutional feature map. Instead
of relying on a selective search algorithm on the feature map for identifying the region
proposals, a separate network is used to predict the region proposals.
Faster R-CNN is faster and can even be used for real-time object detection.

\subsection{Pneumonia Detection Works}
Many researchers have sought to detect pneumonia in the recent past. Abiyev and Ma’ aitah
[15] applied a CNN on the chest x-ray in comparison to RNN the convolutional neural network
gets higher accuracy but has a longer training time.
Guendel [16] proposed to use the DenseNet for the detection on chest x-ray dataset.
Abiyev and Ma’ aitah [15] extracted the features from the layers of CNN and explored them
like GIST on more than 600 radiographs.
As covid pandemic engulfed the world in 2020 Wang and Wong [17] proposed COVID-Net
which is a deep CNN specialized for detection of covid-19 cases from chest x-ray (CXR)
images. Ozturk [18] proposed an automatic detection using raw chest X-ray images; this model
is used to provide accurate diagnostics for binary classification (COVID vs. no findings) and
multiclass classification (COVID vs. no findings vs. pneumonia).

\section{Methodology}
In this part, I would introduce in detail our proposed model method, including data processing,
architecture and enhancement effect used of soft non maxial suppression.

\subsection{Dataset}

The dataset [19] for chest x-ray images and metadata is provided by National institute of health
and clinical research through kaggle. This dataset contains images from 27864 unique patients.
Each image is labelled with three classes.
The case for lung complications is when the lungs are replaced by something other than air like
bacteria, fluids, cells. This causes lung opacities to differ which is why xray beams are used
as such contingent lungs opacity greater than normal because lung tissue is not healthy.
Normal class had images of perfectly healthy lungs without any pathological diseases found in
cxr.
The third class in the dataset lung opacity had images of lungs with clouds associated with
diseases such as pneumonia.
This region of lung opacity is labelled with bounding boxes. An image can have multiple such
boxes If more than one area is detected with more opacity by object detection model. The
middle class are of patients with more opaque lungs but no lung contingencies.

\begin{table}[h]
\centering
\begin{tabular}{|c|c|c|}
\hline
\multicolumn{3}{|c|}{Table 1: Distribution Of Classes in the dataset} \\ \hline
CLASS & Pathogens (1) & None (0) \\ \hline
Lung Opacity & 1 & 9555 \\ \hline
Not Normal / No Opacity & 0 & 11821 \\ \hline
Normal & 0 & 8851 \\ \hline
\end{tabular}
\caption{Distribution Of Classes in the Dataset}
\label{tab:class-distribution}
\end{table}

\subsection{Evaluation Metrics}
The model is evaluated on the mean average precision at different intersections over union IoU
thresholds. The IoU of a set of predicted bounding boxes and ground truth boxes are given by

IoU(A,B) = $\frac{A \cap B}{A \cup B}$

The metric over a range of IoU thresholds at each point calculating an average precision value.
At a threshold of 0.5 a object detected is considered a hit if its IoU with a ground truth object
is greater than 0.5.
At each threshold value t a precision value is calculated based on the number of True positives
(TP), False negatives (FN) and false positives arising from model compared with testing truth
object

A true positive is counted when a single predicted object matches a ground truth with an IoU
above threshold. A false positive is when model predicted object had no associated truth object.
A false negative is when a testing object has no associated predicted object.
When no ground truth objects testing at all for an image, any number of false positives will
result in the image receiving a score of zero and will be included in mean average precision.
The average precision of a single image is calculated as the mean of precision value at each of
IoU threshold.

mAP(p,t) = $\frac{1}{\lvert\text{thresholds}\rvert} \sum_{t} \frac{TP(t)}{TP(t) + FP(t) + FN(t)}$

In this model we are using confidence level for each of bounding boxes. Bounding boxes will
be used to evaluate in order of the confidence levels in the said process.
That is the bounding boxes with higher confidence will be checked first against testing solution
which determines what boxes considered are true and false positives.
None of the edge cases are known to exist in the data set.
Lastly the score returned by the competition metric is the mean taken over the individual average
precision of each image in test dataset.

\subsection{Model}

\begin{figure}[h]
    \centering
    \includegraphics[width=0.7\textwidth]{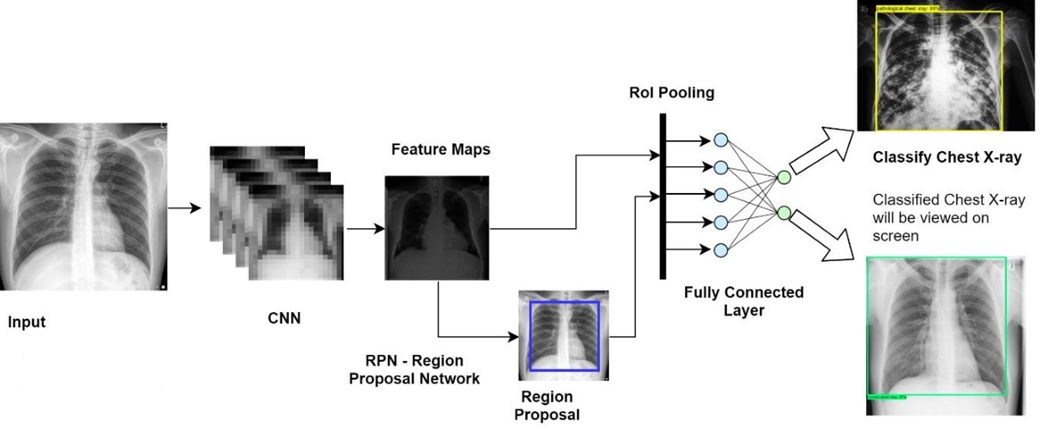}
    \caption{Model}
    \label{fig:model_architecture}
\end{figure}

To avoid test time augmentations which would require high graphical usage and pseudolabelling which is not possible and feasible in practice. To reduce the memory footprint and time of inference. I propose a faster rcnn model utilizing a Resnet encoder which is pertained on TensorFlow ImageNet. Also because Soft-NMS [23] improves the performance of detection
model I have precoded the soft-nms algorithm. 
Faster RCNN is used for object detection while ResNet is the architecture which performs feature extraction for the model.
Faster RCNN defines label for each input, defines how the features [24] are used and label to perform supervised learning, defines the loss function and optimiser and defines training and testing pipeline. While resnet performs how we will extract the feature [25].
After having saved the checkpoints for the trained model we call the model with argument = generatePredictions with the path to model checkpoint and generate predictions for test images.

\section{Implementation}

\subsection{Model Building}
The faster rcnn model is implemented on cloud pytorch with resnet architecture.
The input to the model is a list of tensors of c-color h-height w-width for each image and in
range 0-1. Different images have different sizes
The model behaviour changes depending on if it is training or in evaluation.
During training the model expects both the input tensors as well as targets containing the boxes
which are the ground-truth boxes and labels for each of these ground truth boxes. The model
returns a Dict[Tensor] during tensors containing the classification and regression losses for the
regional proposal network and R-CNN.
During inference the model requires only the input tensors and returns processesed predicted
results as List[Dict[Tensor]], one for each input image.
FasterRCNN needs following arguments as inputs
- a backbone network which is used to compute the features for the model. The backbone
should have a at least an attribute which fives the number of output that each feature
map has. The backbone returns a single Tensor.
- number of output classes for the model. If boxPredictor is specified then number of
classes are none.
- Min and max size of the image to be rescaled before feeding it to the backbone.
- ImageMean the values used for input normalisation. Should be the mean values of
dataset on which the backbone has been trained on.

Mean average precision at differnet intersection over union (IoU) threshold is calculated.
Images with no ground truth bounding boxes are not included in the map score unless there is
a false positive detection.
None is returned if both are empty, don't count the image in final evaluation.
True Positive (TP) tp = tp+1 if matched = True is counted when bt (boxes\_true) are matched.
False Negative (FN) is when the truth box bt has no match then it is calculated as FN fn = fn+1
False Positive (FP) is the box predicted that is not matched by any truth boxes
fp = len(boxes\_pred) – len (matched\_bt)
m = tp/ (tp+fn+fp).
map total += m the score for the mAP is given by map\_total/len (thresholds)
- Images with empty boxes are added to model for contributing to optimisation and loss
calculation
- Original RetinaNet implementation in pytorch [20] ignored images with no boxes while
this model has added them to get better loss calculation and optimisation.
- the small anchors output is added to handle smaller boxes
- to reduce overfitting dropout was added to classification to achieve regularisation and
optimal classification result at same epoch.

The base model preparation part is not that comparatively quicker as PyTorch already provides
a Faster RCNN ResNet50 FPN model with also preTrained implementations so to save time
and computing power, we need to:
Load that model and its comparitive models for results.
Append the model for our input and output needs.
Modify the base model heads with the number of classes according to our dataset.

\subsection{Model Training}
Training dataset included data for 27864 and used data for 1000 participants for testing set.
The range of pretrained pytorch imagenet basemodel dataset was used to pretrain our model.
Without which the model worked on regression but failed on classification.
As training dataset was reasonably balanced there was no need to extra balance. For learning
rate used the ReduceLRonPlateu with patience of 4 and decrease factor of 0.2. Entire image
classification loss and boxes classification were combined to get as total loss.
Figure 8 shows Model Comparision Shows validation losses for a range of various backbones.
The SE-Type nets demonstrated optimal performance with resnext101 showing the best results
and resnet50 being slightly worse. Different architecture were compared with our architecture PNASNet, NASNet, Inception,
SE\_ResNet, Xception and ResNet. There were significant tradeoff between the speed and
complexity and accuracy also parameteres. ResNet demonstration between accuracy and
complexity was optimal. VGG nets did not provide accuracy on dataset while SeNet and
NasNet had the longest training time period.

\begin{figure}[h]
    \centering
    \includegraphics[width=0.7\textwidth]{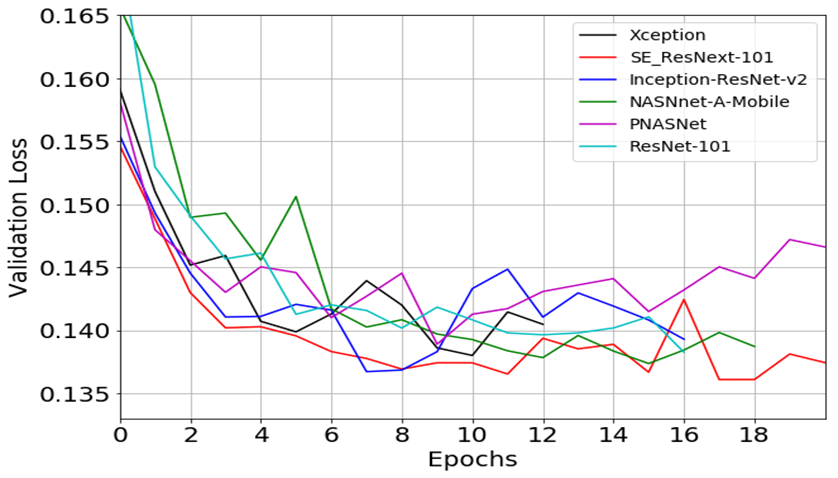}
    \caption{Model Training}
    \label{fig:model_training}
\end{figure}

The Training configuration file contains
- Number of epochs to train for.
- Computation method used for training as the CPU is slow for Faster-RCNN we need
GPU alternatively can use collaboratory by google for object detection in general.
- The batch size used for training
- The dimensions for augmentation we need to resize the images too.

\subsection{Model Learning}
For removal of parameters we trained the model with fixed augmentations and without
classification outputs. Result was improved when making the model predict other related
functions other than only output of regression based model. Without pretraining the model took
much longer for the validation loss to converge pretained in cloud imagenet and dropouts of
0.5 and 0.75 showed the best results on dataset
\begin{figure}[h]
    \centering
    \includegraphics[width=0.7\textwidth]{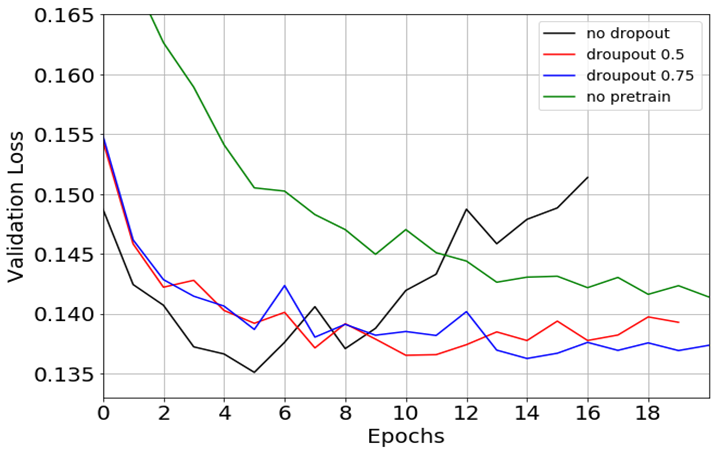}
    \caption{Model Learning}
    \label{fig:model_learning}
\end{figure}

\subsection{Dataset Image Preprocessing}
The dataset was scaled to 512x512 pixels and the 256 pixels gave unsatisfactory results. The
original images were 2000x2000 pixels hence was not practical being heavier to train the
dataset. So deployed following augmentations rotations, shift , horizontal flip. Without enough augmentation the model was overfitted as validation stopped improving on
increasing the training. Heavy augmentation led the model to show better validation loss and
mAP results.
The grayscale is different for the images and low brightness contrast of cxr leads to higher
validation losses as it makes harder to locate the lesion [21] and hence the model proposed in
this paper used clahe algorithm [22] to equal the gray scale of our dataset.

\subsection{Result And Analysis}
The deep learning model is used to perform and detect lung contingencies in chest xray images
dataset.
Faster RCNN + ResNet was trained for 15 epochs.
The Resnet model was compared with Xception, PNASNet And Inception and showed
considerably greater accuracy and specifity for the same dataset used in comparision.

\begin{table}[h]
\centering
\begin{tabular}{|c|c|c|c|c|c|}
\hline
Method   & AC     & SP     & PR     & RC     & F1     \\ \hline
Xception & 93.16\% & 89.03\% & 94.63\% & 87.26\% & 94.5\% \\ \hline
Resnet   & 95.28\% & 96.36\% & 96.6\%  & 97.32\% & 96.93\% \\ \hline
PNASNet  & 94.35\% & 91.79\% & 96.48\% & 89.35\% & 95.74\% \\ \hline
Inception & 95.23\% & 92.64\% & 94.35\% & 87.62\% & 97.63\% \\ \hline
\end{tabular}
\caption{Performance Metrics by Method}
\label{tab:method-performance}
\end{table}

Accuracy, Specificity, Precision, Recall were compared for comparable architecutres as well
as the method used in the paper. The faster rcnn model was trained to classify the cxr dataset.

For evaluation of the validity of the model 5 fold cross validation was performed with first fold
used to test and remaining used to train the model. Then performance as a binary classification
problem.
Performance change in resnet on each fold cross validation shown in table 3
\begin{table}[h]
\centering
\begin{tabular}{|c|c|c|c|c|c|}
\hline
Fold Number & AC    & SP    & PR    & RC    & F1    \\ \hline
Fold-1     & 97.55\% & 97.10\% & 98.0\% & 97.12\% & 97.55\% \\ \hline
Fold-2     & 98.0\% & 96.23\% & 97.0\% & 98.97\% & 97.97\% \\ \hline
Fold-3     & 97.7\% & 97.4\% & 99.1\% & 97.46\% & 98.71\% \\ \hline
Fold-4     & 93.75\% & 96.50\% & 91.0\% & 96.92\% & 93.57\% \\ \hline
Fold-5     & 95.9\% & 96.80\% & 97.0\% & 96.8\% & 96.89\% \\ \hline
Average    & 95.28\% & 96.36\% & 96.60\% & 97.32\% & 96.93\% \\ \hline
\end{tabular}
\caption{Performance Metrics for Folds}
\label{tab:performance}
\end{table}

The Researched FasterRCNN – ResNet Model showcased 95.28
specificity.

\section{Conclusion and Future Scope}
\subsection{Conclusions}
In this paper, faster rcnn model algorithm was used based on resnet architecture. A number of
changes were implemented to improve the accuracy of the model. Also the architecture was
compared with similar models to verify the showcased loss validation results. Heavy
augmentation in particular was applied the dataset. Several checkpoints were utilised in the
model to create a generalise paper. Improvements were made using the said approaches as
shown in the validation loss results.
The model did not involve end user processing and augmentation, gives a good arrangement
between accuracy and speed considering the resources of dataset.

\subsection{Future Scope}
The paper can be compared with other deep learning technologies such as YOLO, SSN and
even add to that different types of CNN like RCNN , Mask RCNN, Fast RCNN
- Different architectures can be used to increase the comparision and verify the paper
results. Even a combination of architectures can be combined to create a newer
architecture which would change the paper labelled tradeoff between the accuracy and
dataset limitation
- The model needs to be trained largerly on a localised GPU for a bigger dataset to verify
that the projected data holds true in further experiments.
- The aspects to solve in further studies involves increasing the learning dataset which
would need labelling done manually by a medical professional and verified by another
medical professional to mitigate any human errors.

\section{Backmatter}

\begin{backmatter}

\bmsection{Acknowledgments}
I would like to express my special thanks of gratitude to my project advisor, Dr. Narendra Singh Yadav, for their able guidance and support. Their help, suggestions, and encouragement greatly contributed to the completion of my report.

I would also like to thank Manipal University and the Department of Information Technology for providing me with all the facilities required for this project.

\bmsection{Data Availability Statement}
The dataset [19] containing chest X-ray images and metadata is provided by the National Institute of Health and Clinical Research through Kaggle.

\bmsection{Data Availability}
Data underlying the results presented in this paper are available in Dataset 1.

\end{backmatter}


\cite{Zhang:14,R2,R3,R4,R5,R5,R6,R7, R8,R13,R14,R17,R20,R21,R22,R9,R12,R15,R18,R23,R10,R11,R16,R19,R24,R25,R26}

\bibliography{sample}

\end{document}